\documentclass{aastex631}
\shorttitle{Data-constrained Simulation of a Solar Jet}
\graphicspath{{./}{figures/}}

\newcommand{\dodoi}[1]{doi: \href{https://doi.org/#1}{#1}}

\begin{document}

\title{Simulation of a Solar Jet Formed from an Untwisting Flux Rope \\Interacting with a Null Point}

\author{Jiahao Zhu}
\affiliation{School of Astronomy and Space Science and Key Laboratory for Modern Astronomy and Astrophysics, Nanjing University, Nanjing 210023, People’s Republic of China; \href{mailto:guoyang@nju.edu.cn}{guoyang@nju.edu.cn}}

\author{Yang Guo}

\affiliation{School of Astronomy and Space Science and Key Laboratory for Modern Astronomy and Astrophysics, Nanjing University, Nanjing 210023, People’s Republic of China; \href{mailto:guoyang@nju.edu.cn}{guoyang@nju.edu.cn}}

\author{Mingde Ding}
\affiliation{School of Astronomy and Space Science and Key Laboratory for Modern Astronomy and Astrophysics, Nanjing University, Nanjing 210023, People’s Republic of China; \href{mailto:guoyang@nju.edu.cn}{guoyang@nju.edu.cn}}

\author{Brigitte Schmieder}
\affiliation{LESIA, Observatoire de Paris, Université PSL, CNRS, Sorbonne Université, Université de Paris, 5 place Jules Janssen,
92190 Meudon, France}
\affiliation{Centre for Mathematical Plasma Astrophysics, Dept. of Mathematics, KU Leuven, 3001 Leuven, Belgium}
\affiliation{University of Glasgow, School of Physics and Astronomy, Glasgow, G128QQ, Scotland}

\begin{abstract}

Coronal jets are eruptions identified by a collimated, sometimes twisted spire. They are small-scale energetic events compared with flares. Using multi-wavelength observations from the Solar Dynamics Observatory/Atmospheric Imaging Assembly (SDO/AIA) and a magnetogram from Hinode/Spectro-Polarimeter (Hinode/SP), we study the formation and evolution of a jet occurring on 2019 March 22 in the active region NOAA 12736. A zero-$\beta$ magnetohydrodynamic (MHD) simulation is conducted to probe the initiation mechanisms and appearance of helical motion during this jet event. As the simulation reveals, there are two pairs of field lines at the jet base, indicating two distinct magnetic structures. One structure outlines a flux rope lying low above the photosphere in the north of a bald patch region and the other structure shows a null point high in the corona in the south. The untwisting motions of the observed flux rope was recovered by adding an anomalous (artificial) resistivity in the simulation. A reconnection occurs at the bald patch in the flux rope structure, which is moving upwards and simultaneously encounters the field lines of the null point structure. The interaction of the two structures results in the jet while the twist of the flux rope is transferred to the jet by the reconnected field lines. The rotational motion of the flux rope is proposed to be an underlying trigger of this process and responsible for helical motions in the jet spire.

\end{abstract}


\section{Introduction} \label{sec:introduction}

    Coronal jets are one of the most common phenomena in the solar atmosphere. They are typically collimated beams with helical motions in a transient process. Modern observations of jets have already been taken through a wide range of wavelengths from X-ray to extreme ultraviolet (EUV) \citep{Shibata_1992,Cirtain_2007,Patsourakos_2008}. Observations with high spatio-temporal resolutions from the Atmospheric Imaging Assembly \citep[AIA;][]{lemen_2012} on board the Solar Dynamics Observatory \citep[SDO;][]{SDO2012} and simultaneous spectral observations from the Interface Region Imaging Spectrograph \citep[IRIS;][]{Pontieu_2014} provide an opportunity to study the origin and development of jets \citep{Joshi_2020a,Yang_2020}. Although the origins and morphology of jets have been investigated for decades, the results are still inconclusive \citep{LiuC_2011,Shibata_1994,Raouafi_2016,Schmieder_2022a}. Usually jets can be classified into standard jets, which fit the standard reconnection picture, and blowout jets, analogous to a miniature version of blowout explosion at their bases \citep{Moore_2010}. Coronal jets extend in length from tens of Mm to about $400$ Mm and possess a speed from about $100$ km s$^{-1}$ to $600$ km s$^{-1}$ \citep{Nistico_2009,Paraschiv_2010}. They are not isolated phenomena and are believed to be closely related to other activities like coronal mass ejections, particle acceleration in the solar wind, and even to coronal heating and so on \citep{Nitta_2006,Yu_2014,Joshi_2020b}. There are still many questions to be solved for solar coronal jets.
    
    Magnetic reconnection is assumed to be responsible for explosive jet events in the corona. Fan-spine structure is composed of a fan-like dome with a null point connecting an outer spine to a remote region \citep{Parnell_1996,Priest_1996}. The reconnection can easily occur at a null point and transfer the plasma to the far end of the spine. When reconnection takes place at a null point, energy stored in the magnetic field is continuously converted into the kinetic energy and thermal energy of plasma. And then triggered by a breakout structure, an intensive explosion or rapid motion of plasma can be observed in the corona. In terms of the magnetic topology, fan-spine structures \citep{Liu_2011}, eruption of magnetic flux ropes \citep{Adams_2014,Zhu_2017} and bald patches with a continuous build-up of electric current \citep{Guo_2013,Schmieder_2013} may easily trigger jet events. Bald patch is a region where the magnetic field lines are tangential to the photosphere and the direction of field lines points from negative polarity to positive polarity \citep{Titov_1993}. The magnetic configuration of parasitic polarities situated in a large opposite polarity tends to generate such topologies. In some studies, successive jets in coronal holes exhibit a homologous behaviour of recurrence, resulting from impulsive reconnection between closed and open field lines \citep{Pariat_2010}.
    
    Although spectral and EUV images have been widely used in the analysis of the jet formation and evolution, the lack of simulation is still a restriction on exploring the nature of jets \citep{Schmieder_2022a}. Recently, state-of-the-art magnetohydrodynamic (MHD) models with high performance computing techniques help to understand the origin of jets. \citet{Fang_2014} simulated the process of a twisted flux rope emerging in an open magnetic field and found that magnetic reconnection is responsible for the explosive events. \citet{Gonz_2020} performed simulations of different parameters to probe the impacts on the origin and morphology of the jets. They found the significance of the resistivity and thermal conductivity in constructing the jets. \citet{Wyper_2019} simulated a helical jet caused by a sigmoidal flux rope eruption, with artificial surface motions injecting free energy into the active region. This model exhibits a coupled mechanism of the breakout reconnection and ideal MHD instability in the process of jet eruption. However, most of the simulations about jets mentioned above are ideal MHD models controlled by artificial settings. Simulations utilizing magnetograms from observations have not yet been widely studied. A force-free field is widely used to study the stable magnetic configuration before eruptions \citep{Wiegelmann_2006,Wiegelmann_2021}. Such fields can be described by $\left( \nabla \times B \right) \times B = 0$ and it becomes  $\nabla \times B=\alpha \left( r \right)B$ if considering the divergence-free condition. When the torsional parameter $\alpha \left( r \right)$ equals to zero or is a constant, the field is a potential field or a linear force-free field, which can be solved in such particular situations. To further explore the actual physical conditions, a non-linear force-free field (NLFFF) is solved where $\alpha \left( r \right)$ varies with position. Although no general analytic solutions exist because of the nonlinearity of the equation, various algorithms have been proposed and implemented to solve this issue \citep{Schrijver_2006,Metcalf_2008}. In this aspect, an NLFFF extrapolated from the observed magnetogram and the data-driven simulation based on such a field further help us to understand the magnetic structures of the active region and nature of flux rope eruptions \citep{Guo_2019} including some complex solar eruptions \citep{Jiang_2022}. \citet{Jiang_2016} simulated a coronal magnetic field which transformed from the pre-eruptive to eruptive state following a long-duration quasi-static evolution, and found the establishment of a positive feedback between emerging flux and external magnetic reconnection. Therefore, a promising method is to perform the data-constrained or data-driven simulation that can reproduce the jet eruption with helical motions as observed.
    
    Recently, statistical studies on the origin of coronal jets draw a conclusion that mini-filament eruptions play a critical role in triggering both standard and blowout jets \citep{Sterling_2022,Baikie_2022}. The rotational motion of a mini-filament along with its upward motion to the top of an arch system can result in magnetic reconnection. The reconnection between the inner and ambient field lines decreases the bound force imposed on the mini-filament and thus leads to its eruption. The jet followed with the mini-filament eruption shows a mixture of both cold and hot material and exhibits a helical motion, often seen in coronal mass ejections \citep{Sheeley_2007} and active region flares \citep{Zhang_2022}. These characteristics suggest that the relations between solar eruptions at different scales show quite similar features. Besides, ideal experiments of magnetic reconnection have also been performed. \citet{Farid_2022} aimed to check the presence of a mini-filament. They used the flux-rope insertion method to create an energized NLFFF model of coronal jets and identified the consequent eruption as an untwisting jet. \citet{Pariat_2015} demonstrated that for getting a helical jet, it is a primary requirement to store sufficient magnetic free energy and impulsively release them. They thought the pre-eruption magnetic configuration might not be so important. \citet{Wyper_2017} on the other hand proposed that the magnetic breakout is a universal model for solar eruptions including coronal jets. However, current simulations can not spontaneously produce a mini-filament structure directly from observed magnetograms.
    
    In this paper, we analyze the motion of a jet through multi-wavelength observations from AIA and IRIS. Then we simulate the active region with a zero-$\beta$ MHD simulation using the Message Passing Interface Adaptive Mesh Refinement Versatile Advection Code \citep[MPI-AMRVAC;][]{Keppens_2003,Keppens_2012,Porth_2014,Xia_2018} and find the structure of a null point and a flux rope at the jet base. In Section \ref{sec:observations and method}, we show the observations of the jet and depict its kinematic evolution. The extrapolation and simulation methods are also described. Section \ref{sec:results} shows the results and Section \ref{sec:summary} presents a summary and discussion about our simulation. We show the importance of the twisting flux rope in the reconnection of this jet event.

\section{Observations and Methods} \label{sec:observations and method}

\subsection{Instruments} \label{sec:instruments}

 SDO/AIA provides regular full-disk EUV observations at several different wavelengths with a spatial resolution of $0.6''$ and a temporal cadence of $12$ s. The AIA channels respond to a wide range of temperature, (0.1--10 MK), and can be used to explore solar features over a wide range of scales.

IRIS is widely used to detect the transition region between the chromosphere and corona. It switches to different raster modes and obtains spectral data with different resolutions. Also, it can provide high-resolution slit-jaw images (SJIs) in several wavebands such as C II 1330 \AA, as a supplement to the SDO/AIA observations. The C II 1330 \AA $\ $SJIs cover a specific region with a field of view of $60''  \times 68''$ and a pixel size of $0.16'' \times 0.32''$. We also adopt simultaneous spectroscopic observations right above the reconnection site to analyze the Doppler shift in this region. We use the Mg II k lines in this work, which forms at chromospheric temperatures, and the SJI 1330 for the comparison of the structures of AIA.

\begin{figure}  
    \centering
    \includegraphics[width=15cm]{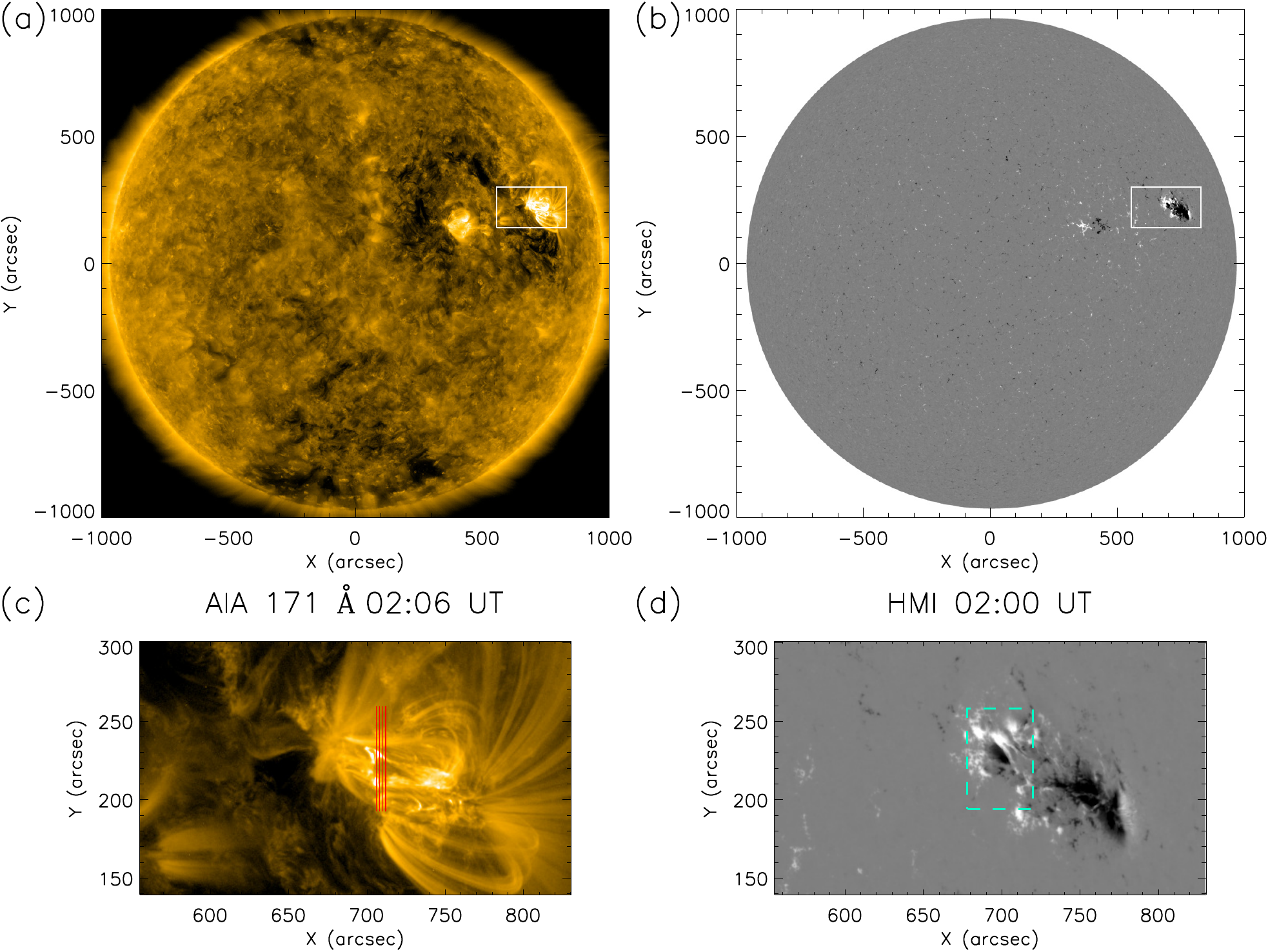}
\caption{(a) Overview of the jet event at 02:03 UT on 2019 March 22 by SDO/AIA 171 \AA. (b) Full-disk magnetogram observed by SDO/HMI from line of sight at 02:00 UT, plotted in limits [-200,200] Gauss to enhance the image contrast. The white squares in (a) and (b) outline the observation region scanned by Hinode/SP. (c) The enlarged view of the white square shown in (a) and the four red lines correspond to the slit positions of IRIS at the corresponding time. (d) An enlarged view of the HMI magnetogram in the white square shown in (b), plotted in limits [-1000,1000] Gauss.} The cyan dashed square is similar to that in Figure \ref{fig:fig5} regardless of projection effects.
    \label{fig:fig1}
\end{figure}

The magnetogram used in the magnetic field extrapolation and numerical simulation comes from the Spectro-Polarimeter \citep[SP;][]{SP2013} of the Solar Optical Telescope \citep[SOT;][]{SOT2008-1,SOT2008-2,SOT2008-3,SOT2008-4} on board Hinode \citep{Hinode_2007}. It was obtained by scanning this region step-by-step along the east–west direction in a fixed band of wavelengths centered on the Zeeman-sensitive Fe I lines at 6302 \AA. The observation started at 01:02:05 UT and ended at 01:59:30 UT, right before the jet event which started at around 02:03 UT. The whole magnetogram covers a field of view of $270'' \times 164''$ with a pixel size of $0.32''$  along the slit and a scanning step size of $0.3''$. To compensate the stretch caused by the long-time scanning, we scale its spatial resolution with the assistance of the magnetogram from the Helioseismic and Magnetic Imager \citep[HMI;][]{Scherrer_2012,Schou_2012} on board SDO by comparing the main common features in the magnetograms, which have already been corrected for projection effects. The pixel size of the HMI magnetogram is $0.5''$. The position information between the instruments we used is shown in Figure \ref{fig:fig1}.

\subsection{Observations} \label{sec:observations}

\begin{figure} 
    \centering
    \includegraphics[width=15cm]{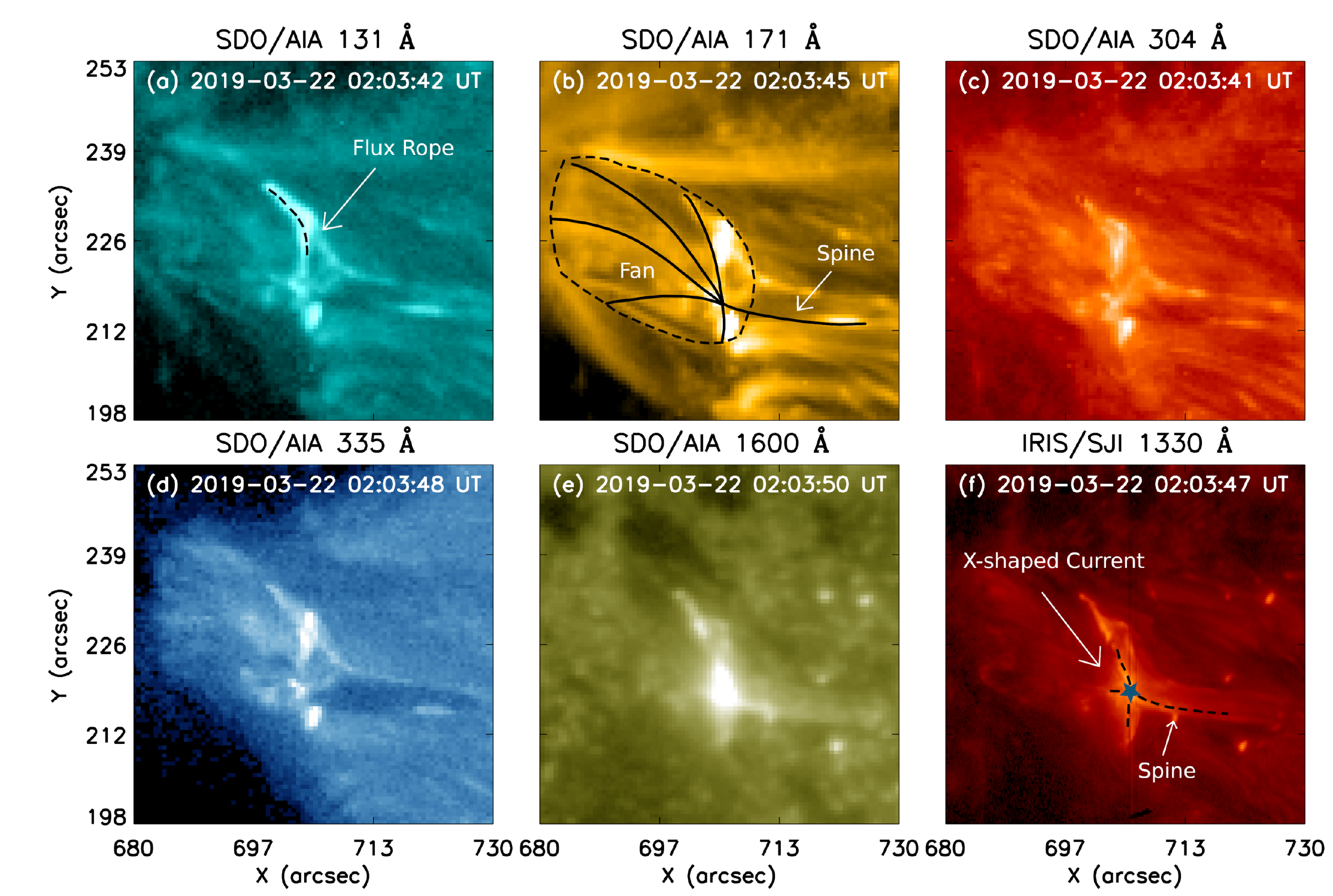}
    \caption{Multi-wavelength observations with (a)--(e) SDO/AIA and with (f) IRIS of the jet at around 02:04 UT on 2019 March 22. The triangular jet base can be clearly observed in all these wavelengths. The flux rope can be figured out by the brightening structure in the north as shown in (a) and the X-shaped current can be recognized in the south as shown in (f). The fan-spine structure is outlined in (b), where the dashed line outlines the base shape of the fan, and the solid lines outline the spine and several field lines constituting the fan. The blue star in (f) represents the position of the null point, where the spine connects the fan to the west remote region. The jet event lasts for several minutes. A movie of the multi-wavelength observations is attached to this figure. The movie traces the jet base on 2019 March 22 from 01:58 UT to 02:14 UT in all the wavelengths presented.}
    \label{fig:fig2}
\end{figure}

    The jet event started at around 02:03 UT on 2019 March 22 in the active region (AR) 12736 and lasted for several minutes. This event has been intensively studied from an observational point of view \citep{Yang_2020,Joshi_2020a,Joshi_2021a,Joshi_2021b,Schmieder_2022a,Schmieder_2022b}. Multi-waveband observations from SDO/AIA recorded this intense eruption in all its wavebands. We select four EUV images and one UV image provided by SDO/AIA as shown in Figure \ref{fig:fig2}, which reflect different temperatures in the chromosphere and corona. At the EUV brightness peak time around 02:06 UT, the helical motion of the jet around its axis can be recognized in most of the wavebands through the bright and dark patterns. Meanwhile, IRIS observed the active region and one of its slit position was just above the reconnection site of the jet base at the early eruption phase. Figure \ref{fig:fig2} shows some observational features, such as the shape of the jet base. An X-shaped reconnection site and a flux rope can also be inferred by the heated brightening structures as shown in Figure \ref{fig:fig2}. SOT/SP on board Hinode obtained magnetic field on the photosphere from 01:02 UT to 01:59 UT by scanning this region with a spectrograph. A potential field can be extrapolated when the magnetic field is in a stable or quasi-stable state. Usually it is valid doing so before eruptions occur. As the magnetogram shows, this magnetic field possesses a major positive polarity at northeast and a major negative polarity at southwest, and several parasitic negative polarities around the positive one. A fan-spine structure may easily form in such a magnetic configuration around those parasitic polarities. In this case, the fan lies above main polarities in the east and the outer spine connects to the far side in the west region.

\begin{figure} 
    \centering
    \includegraphics[width=12cm]{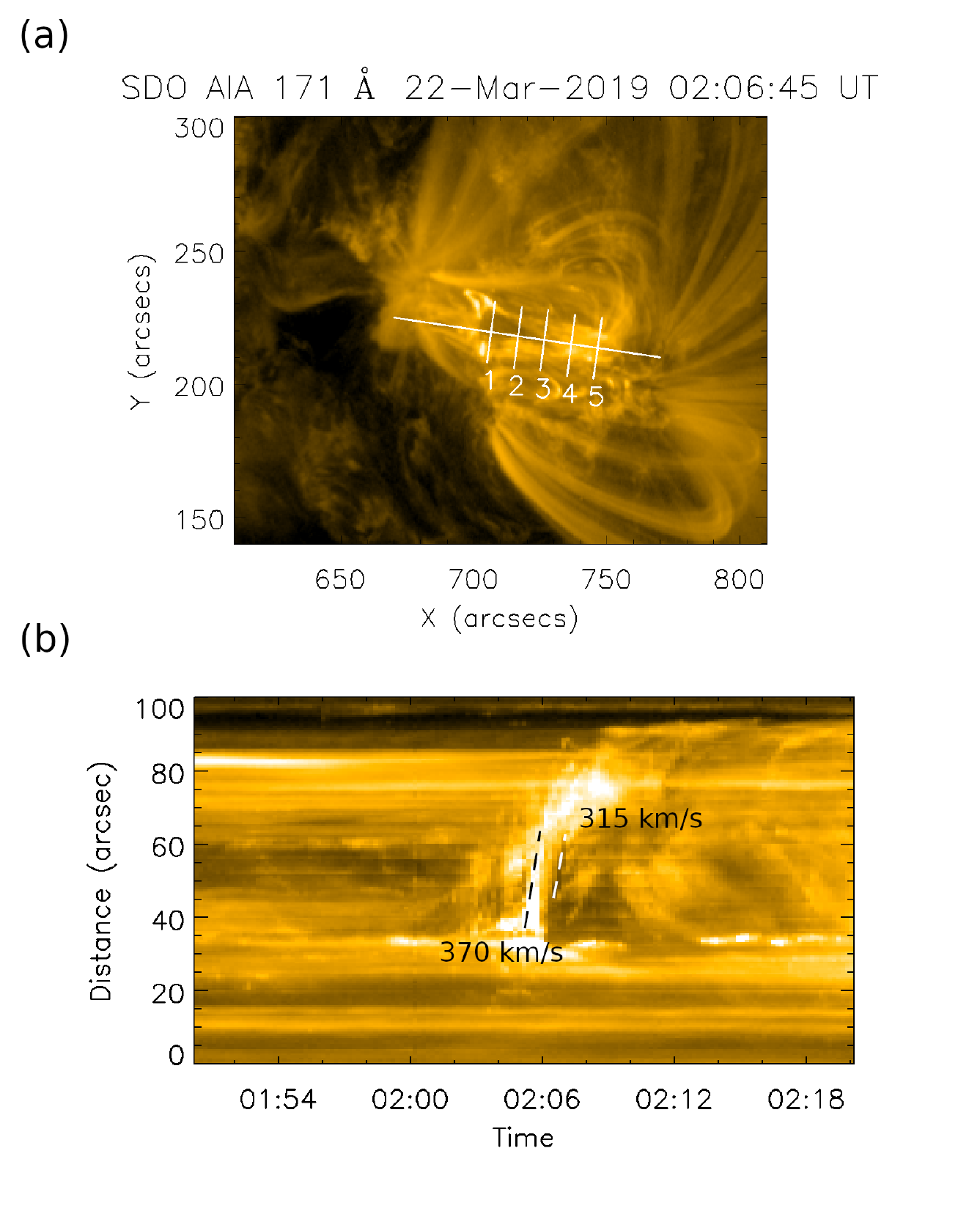}
    \caption{(a) SDO/AIA 171 \AA \ image at 2:06 UT on 2019 March 22. The long white line represents the slice used in the time-distance analysis, which is mainly along the jet propagation direction. The five short slices perpendicular to the long one are used to reveal the untwisting motion of the flux rope. (b) Time-distance map of the slice along the jet propagation direction. The length of this slice is $100''$. The bright pattern shows eruption of hot plasma at a speed of 370 km\ s$^{-1}$ and the dark pattern shows eruption of cold plasma at a speed of 315 km\ s$^{-1}$.}
    \label{fig:fig3}
\end{figure}

\begin{figure}  
    \centering
    \includegraphics[width=8cm]{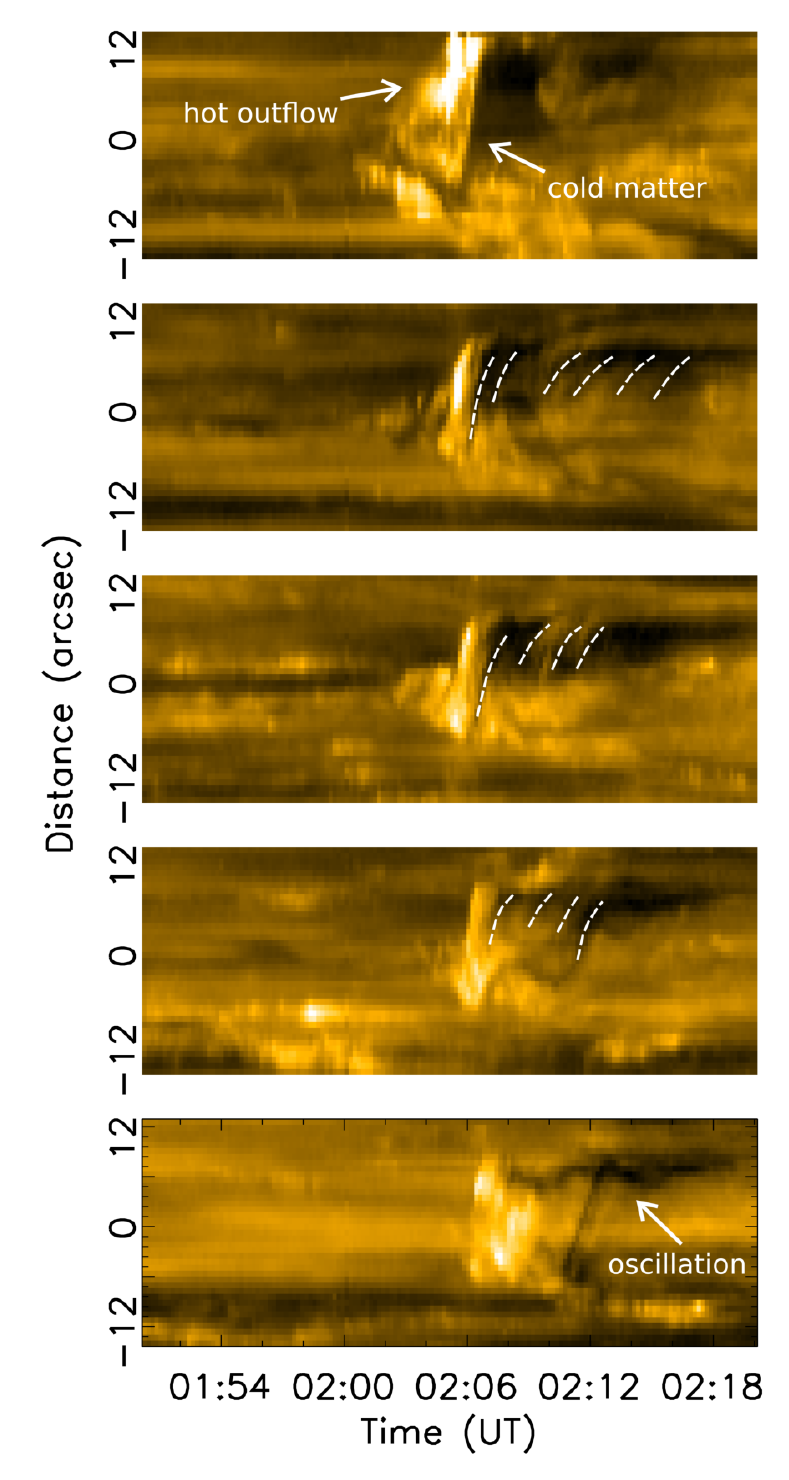}
\caption{Time-distance maps of the five short slices perpendicular to the jet propagation direction. The panels from top to bottom correspond to the slices 1 to 5 in Figure \ref{fig:fig3}. The length of these slices is $24''$ and the spacing is $10''$. The white dashed lines indicate the helical motion of the jet event perpendicular to the spire. These maps also reveal a transverse oscillation of the jet and the mixture of hot and cold material propagating separately.}
    \label{fig:fig4}
\end{figure}

We focus on the small field of view of the jet base using the SDO/AIA full-disk observations. To study the propagation of the twisting jet, we select a slit along the direction of the jet indicated as the solid white line in Figure \ref{fig:fig3}(a) and present the time-distance map of the slit in Figure \ref{fig:fig3}(b). The figure shows the evolution of the jet spire, a bulk motion of bright plasma with a velocity of $\sim 300$ km\ s$^{-1}$ followed by a motion of dark structure at around 02:04 UT. This suggests that the hot outflows set in through magnetic reconnection at first, followed by the eruption of the cold material, which may come from the flux rope in the jet base, and move along the reconnected field lines. The curved pattern in Figure \ref{fig:fig3}(b) also shows the helical motion of the jet, indicating that a twist structure, probably a flux rope, may exist previously at the jet base. To further look into the helical motion of the jet, we draw five equi-distanced cuts perpendicular to the slit in Figure \ref{fig:fig3}(a). Figure \ref{fig:fig4} displays the time-distance maps of the five cuts. The eruptive motion of the bright and dark patterns shown by slices in two perpendicular directions in both Figure \ref{fig:fig3}(b) and Figure \ref{fig:fig4}, as well as the recurrent features labeled by dashed white lines in Figure \ref{fig:fig4}, exhibit the helical motion of the jet. The panels of Figure \ref{fig:fig4} in sequence also shows the separation between these two patterns. The dark wave feature in the last panel of Figure \ref{fig:fig4}, which is sliced from the transverse direction, clearly shows an oscillatory motion in the eruption.
	
IRIS performed spectral observations with a 4-step raster for the region covering the reconnection site during the whole process of the jet event. At the time of the jet eruption, the profile of the Mg II lines shows a broad extension in both the red and blue wings, which indicates complex material motions in the transition region. The presence of complex motions has been separately verified by \citet{Joshi_2020a}. We adopt the moment method to calculate the average Doppler velocity regardless of the irregular profile. We mainly analyze the Mg II k lines of the first slit position, which is located just above the reconnection site during the peak time. We thus obtain the Doppler velocity from observations around the peak time and then compare it with the numerical values calculated from the data-constrained simulation in Section \ref{sec:results}.

\subsection{Magnetic field extrapolation and data-constrained simulation} \label{sec:extrapolation}
Based on the vector magnetic field obtained from SOT/SP, we make an NLFFF extrapolation to study the magnetic topology structure in AR 12736. To further investigate the relationship between the magnetic structure and observational morphology, we scale the spatial resolution of the SOT/SP magnetogram by comparing it with the SDO/HMI magnetogram that has already been corrected for the projection effects. Meanwhile, we resize the original data to reduce the spatial resolution by twice for relieving the computational burden. The NLFFF is reconstructed from the magneto-frictional module implemented by \citet{Guo_2016a,Guo_2016b} in MPI-AMRVAC. It can relax the potential field together with the bottom boundary to NLFFF in the computation domain. The method can be applied to multiple situations, such as Cartesian or spherical coordinates with uniform grids or adaptive mesh refinement grids. We note that it is hard to attain a completely force-free field relaxed from an observed magnetogram, because the Lorentz force deduced from observations is not nicely balanced in the photosphere due to the role of gravity \citep{Zhu_2016}. So, the NLFFF might not be in a complete equilibrium state initially.

\begin{figure}  
    \centering
    \includegraphics[width=14cm]{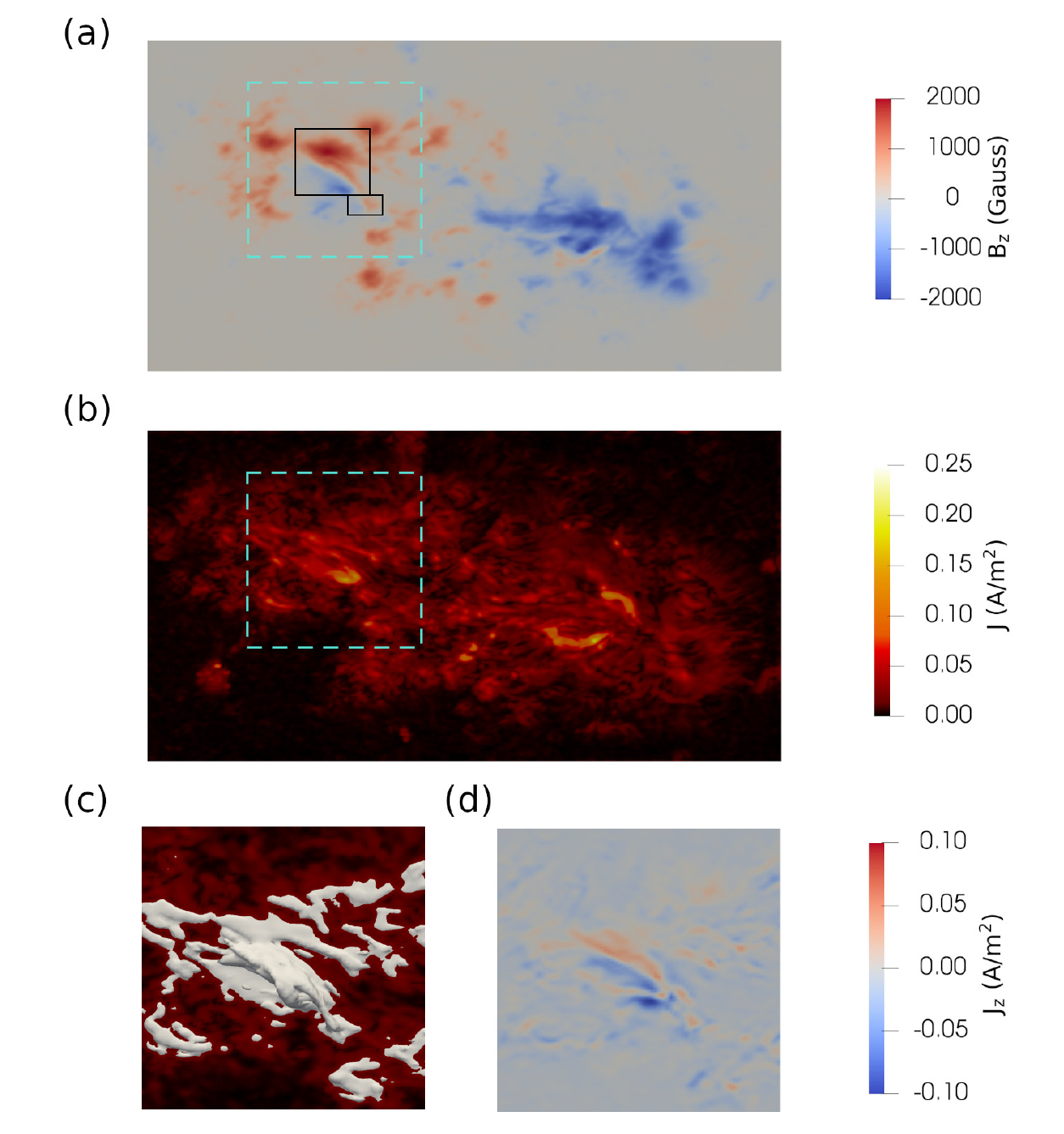}
    \caption{(a) Vertical magnetic field on the layer of 1.2 Mm above photosphere in the NLFFF extrapolation. The cyan squares in panels (a) and (b) are also labeled in Figure \ref{fig:fig1}(d) regardless of projection effects and outline the field of view in panels (c) and (d) where current mainly distributes. The black rectangles show the main polarity and parasitic polarity in the region of interest. (b) Distribution of the current density magnitude $J$ on the same layer. (c) An iso-surface of 0.03 A m$^{-2}$ of $J$ overlaid on the current density distribution. The concentrated current shown by the iso-surface indicates the flux rope structure. (d) Distribution of $J_{z}$ on the same layer as that in panel (a). There exists an elongated double-peaked $J_{z}$ pattern, which reveals a long flux rope structure along the polarity inversion line. The field of view has been corrected from the projection effects, therefore the structures are more elongated along the $x$ axis compared to Figure \ref{fig:fig1}.}
    \label{fig:fig5}
\end{figure}

\begin{figure}  
    \centering
    \includegraphics[width=14cm]{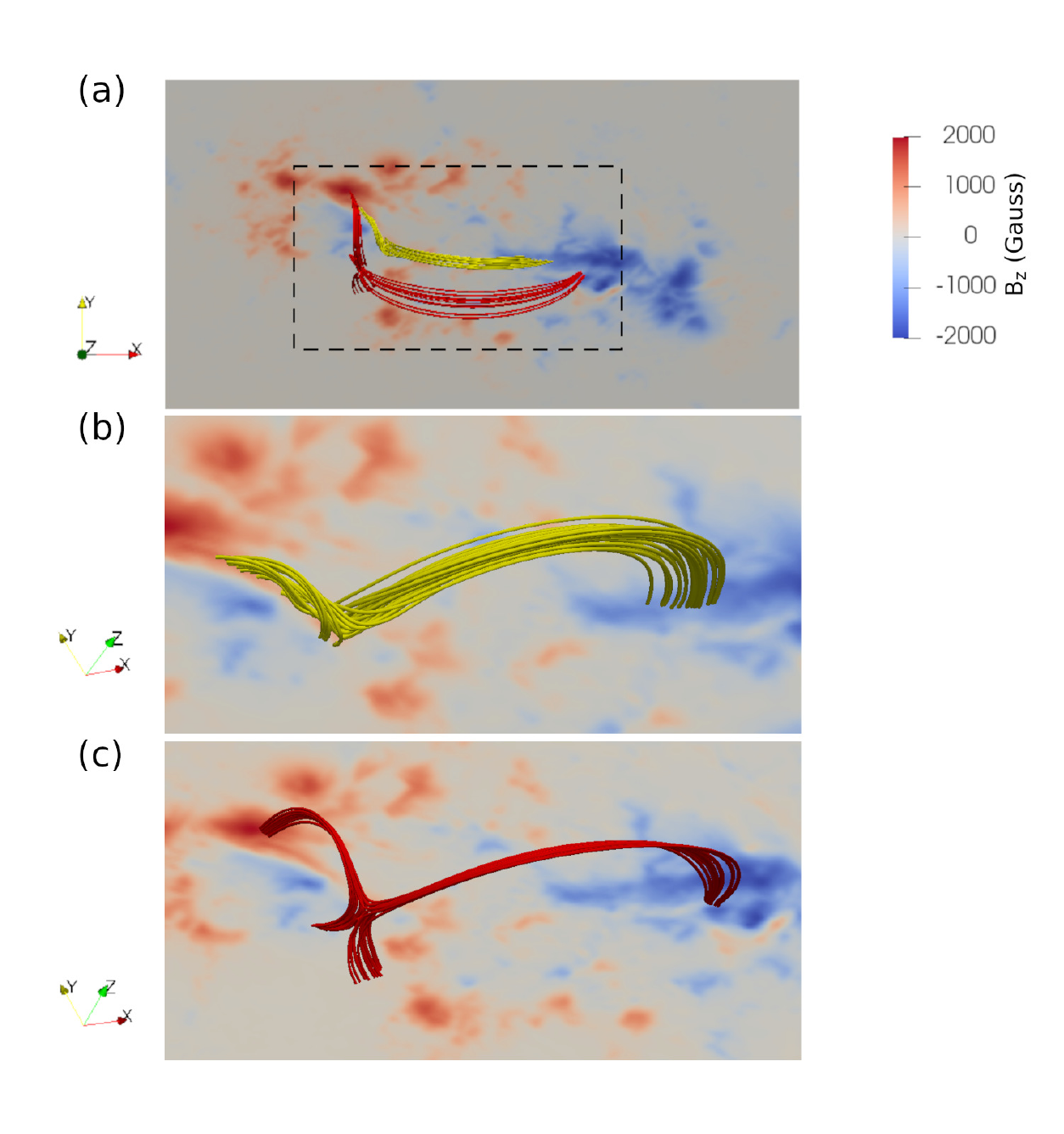}
    \caption{(a) Sample magnetic field lines in the NLFFF model showing different magnetic structures. The background shows the vertical magnetic field $B_z$. The rectangle region contains two main structures and is enlarged in panels (b) and (c) with a side view. The orientations are labeled by the orientation axes in the small left panels. (b) A flux rope structure colored by yellow field lines lying above a bald patch near the photosphere in the north region. (c) A null point structure in the X-shaped current sheet colored by red field lines in the south region. }
    \label{fig:fig6}
\end{figure}

Figures \ref{fig:fig5}(a) and \ref{fig:fig5}(b) show the vertical magnetic field and the distribution of current density magnitude in the extrapolated field 1.2 Mm above the photosphere. We zoom in the region of interest showing a current density iso-surface of 0.03 A m$^{-2}$ and the distribution of $J_{z}$ in Figures \ref{fig:fig5}(c) and \ref{fig:fig5}(d), respectively. The concentrated current reveals a strong magnetic flux rope structure in the vicinity of the small parasitic negative polarities in the north. It is the same structure detected in the HMI vector magnetograms by \citet{Joshi_2020a}. And the elongated double-peaked $J_{z}$ pattern with hooks confirms the existence of the flux rope \citep{Aulanier_2019,Joshi_2020a}. Figure \ref{fig:fig6}(a) displays two magnetic structures in the extrapolated field, a flux rope and a null point. Figure \ref{fig:fig6}(b) shows the flux rope lying above a bald patch region low in the photosphere. By tracing the magnetic field lines of the flux rope, we find that they are actually a part of the bald patch field lines. Figure \ref{fig:fig6}(c) shows the null point in the south above the flux rope. The X-shaped reconnection site is reproduced with the spine connecting to a remote region in the west.

To probe the nature of the jet, we conduct a data-constrained MHD simulation based on the result of the NLFFF extrapolation. The model we adopt is the same as that in \citet{Guo_2019}, which solves the zero-$\beta$ MHD equations. For the boundary conditions, we select the data-constrained case, which fixes the bottom boundary, rather than the data-driven case, because of the limitation of the scanned magnetogram from SOT/SP. The MHD equations are solved in the dimensionless forms in MPI-AMRVAC. In this model, the gravity and gas pressure gradient are omitted in the momentum equation, and the energy equation including the heat conduction term and others is dropped entirely. In other words, we only solve the continuity equation, momentum equation and magnetic induction equation to get the density, velocity, and the magnetic field.

The initial condition is set to be the relaxed NLFFF and the initial density distribution is calculated by assuming a temperature distribution of a stepwise function of height. We set the computation grid to be $288 \times 200 \times 200$, namely $166 \times 85 \times 85$ Mm$^3$ in real scale with the resolution to be $0.58 \times 0.42 \times 0.42$ Mm$^3$. Also, an anomalous resistivity is added in this model to facilitate the occurrence of magnetic reconnection. When the current density $J$ of a local point is higher than the critical value $J_{c}$, the resistivity $\eta$ is set to be $\eta = \eta_{0} \times ({J-J_{c}})^{2}/{J_{c}}^{2}$ and otherwise to be zero. Here, we set $\eta_{0}$ and $J_{c}$ to be $5 \times 10^{-4}$ and $1000$ in normalized units, respectively. The whole simulation time is about half an hour. We output the snapshots of the simulation results every 24 s for further comparisons with the AIA and IRIS observations.

\section{Results} \label{sec:results}

\begin{figure}  
    \centering
    \includegraphics[width=18cm]{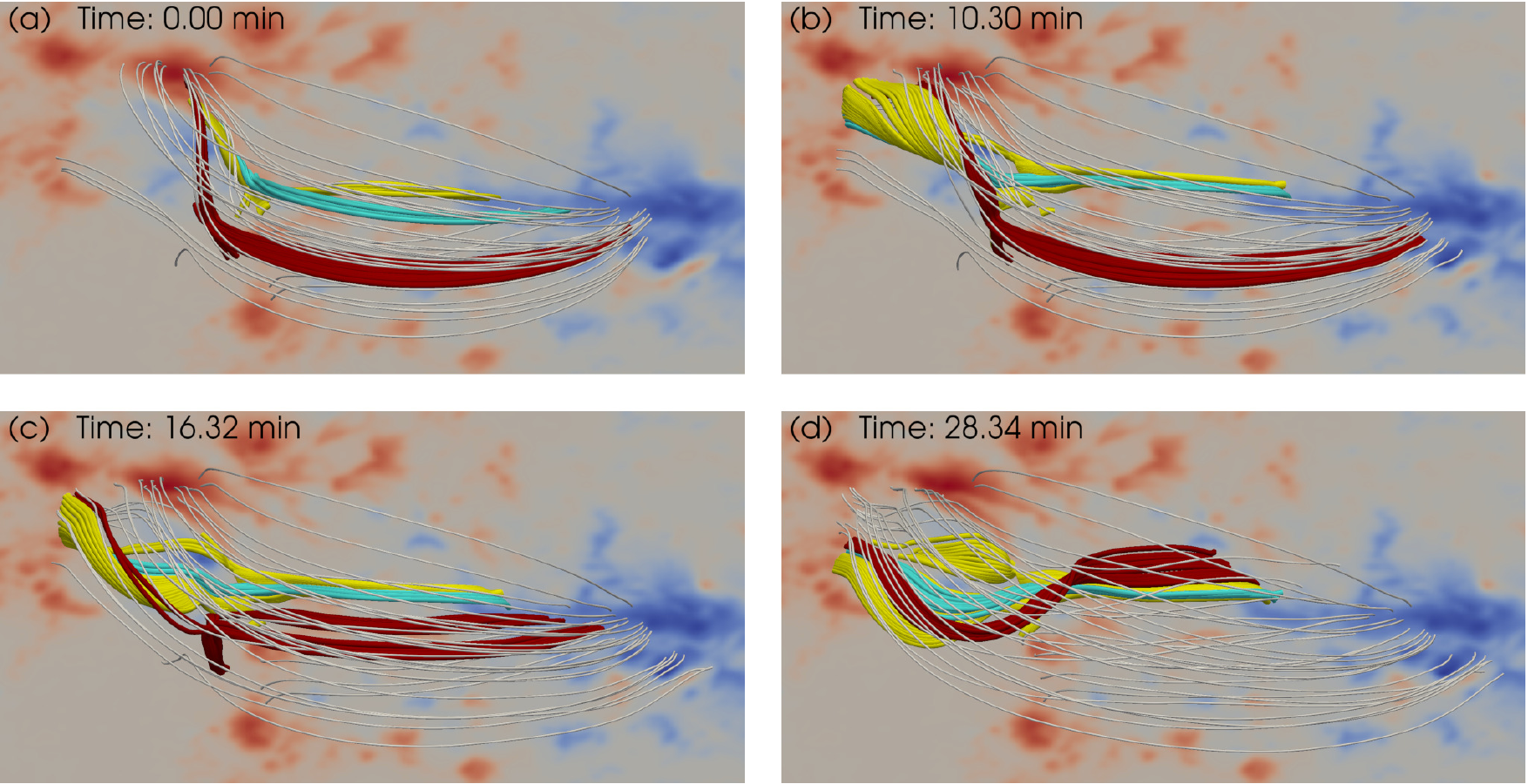}
    \caption{Magnetic field evolution in the zero-$\beta$ simulation. (a)--(d) Four moments of the simulation. Red field lines represent the null point structure and the reconnection site. Cyan and yellow field lines depict the flux rope structure and gray ones represent the ambient field lines. (a) The initial magnetic field configuration, which is the same as that in Figures \ref{fig:fig6}(b) and \ref{fig:fig6}(c). (b)--(c) The flux rope in the north region slowly moves southwards, rotates counterclockwise and interacts with the field lines connected to the null point. The reconnection outflow and flux rope material move along the long spine of null point connected to a remote region in the west. (d) Magnetic field lines after the rotation of the flux rope and interactions between the flux rope and X-shaped current. The reconnection process ends at this moment. A movie of the data-constrained MHD simulation is attached to this figure. The movie shows the evolution of the simulation in 32 minutes from the local top view.}
    \label{fig:fig7}
\end{figure}

 In the NLFFF extrapolation, we recognize a flux rope low in the photosphere related to the north parasitic polarities, and an X-shaped current sheet high in the corona in a south region between the two main polarities, which is also the reconnection site as shown in Figure \ref{fig:fig6}. At the early time of the simulation, the eastern footpoint of the flux rope slowly drifts close to the positive polarity in its southeast. Meanwhile, the main body of the flux rope rotates around its axis counterclockwise and approaches the reconnection site located high in the south region. The rotation slowly changes the magnetic configuration and the kinetic energy of plasma is accumulated through the continuous reconnection. When the field lines of the flux rope encounter the X-shaped current sheet, they join the reconnection process with other field lines within the sheet. Then, the explosive burst takes place and the jet with helical motion comes out of the null point along the spine connecting to the faraway side in the major negative polarity. At last, the field lines shrink to a low magnetic arch structure after the burst. Thus, we can detect the hot reconnection outflows and later the cold material from the flux rope, corresponding to the bright and dark patterns in AIA observations, respectively. The general evolution is revealed by four typical moments shown in Figure \ref{fig:fig7}.

\begin{figure}  
    \centering
    \includegraphics[width=18cm]{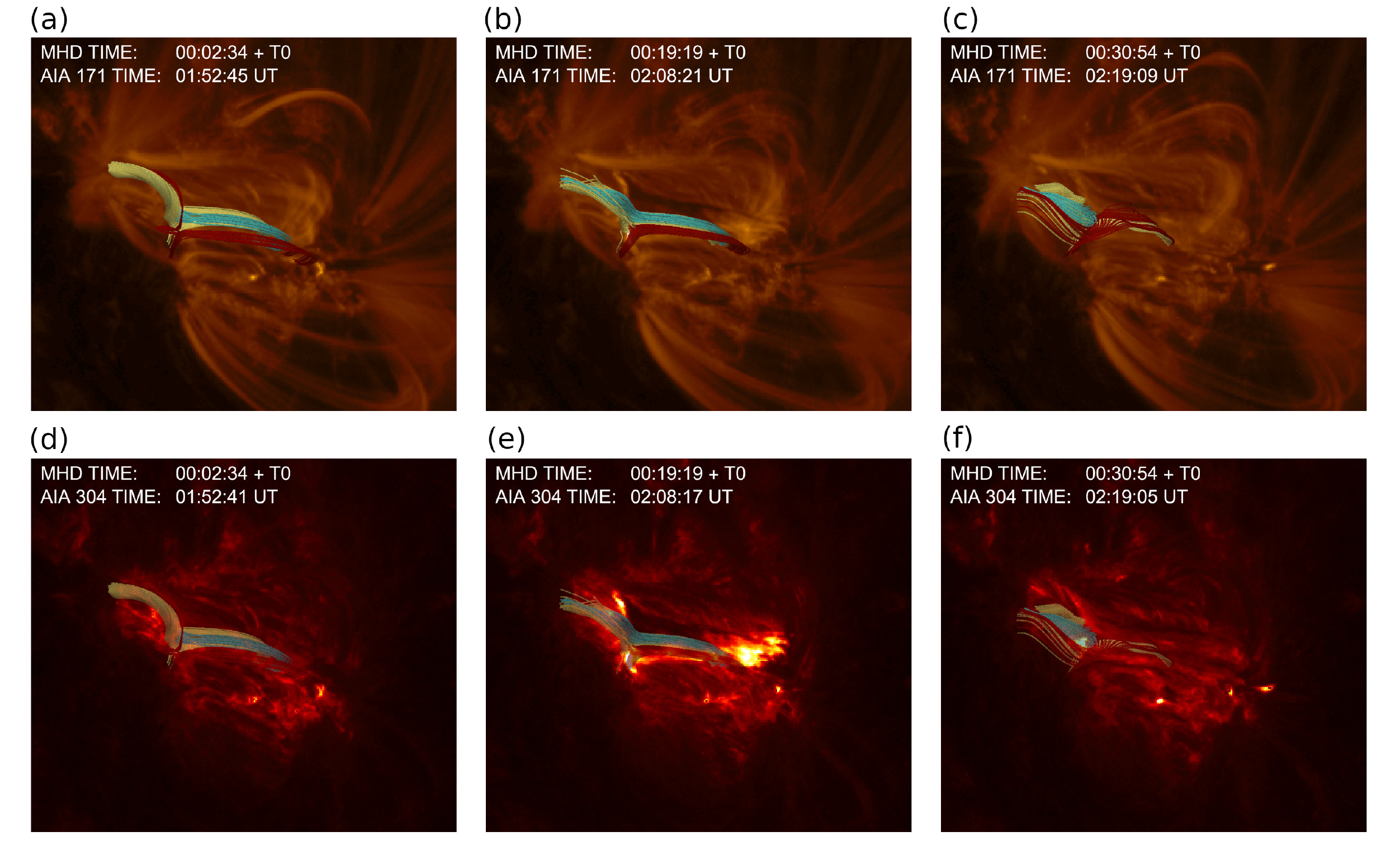}
    \caption{(a)--(c) Simulation results superposed on the SDO/AIA 171 \AA $\ $ images from the viewing angle of the observing instruments at three moments. (d)--(f) Simulation results superposed on the SDO/AIA 304 \AA $\ $ images from the viewing angle of the observing instruments at three moments. The simulation time is expressed in the format as AIA and a T$_{0}$ is used to represent the start time of the simulation. The field lines in our simulation fit well with the arcade and the triangular shape of the jet in observations. In general, the simulation results show an evolution roughly synchronously with the observation. A movie of the results is attached to this figure. The movie shows the evolution of the simulation in 32 minutes superposed on simultaneous SDO/AIA 171 \AA $\ $ and 304 \AA $\ $ observations from the line-of-sight view.}
    \label{fig:fig8}
\end{figure}

To compare the simulation with the observations of SDO/AIA and IRIS, we project the simulation to the line of sight and superpose it on the AIA observations in Figure \ref{fig:fig8}. Since the magnetogram from SOT/SP is scanned through a period of time, we could not give an exact time that the simulation starts from. Thus we use T$_{0}$ to represent the start time and focus on the time evolution of simulations and observations. The simulation matches the observations quite well in the perspective of the triangular shape at the base. The field lines of the flux rope fit with the anemone-like arcade and the field lines at the null point display a triangular shape like the jet. Note that only magnetic field lines are displayed in Figure \ref{fig:fig8}, and the plasma is assumed to move along the field lines. We find that the time-sequential images match roughly synchronously with the real-time evolution revealed in the observations. All these results indicate that the simulation reproduces the observations to a satisfactory degree.

\begin{figure}  
    \centering
    \includegraphics[width=18cm]{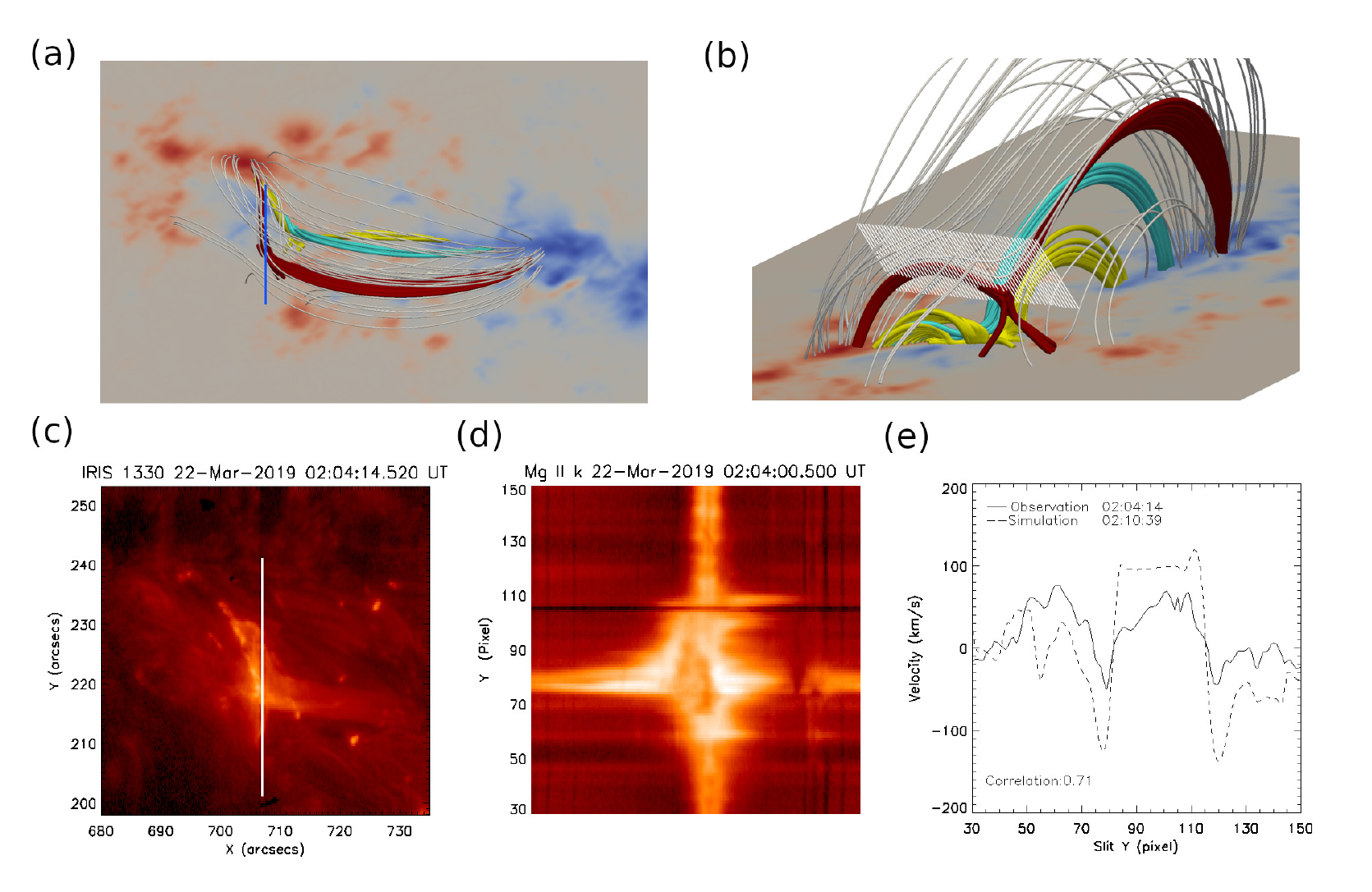}
    \caption{Doppler velocities derived from the observations and simulations. (a) Slit position shown in the simulation at a specific time corresponding to that of observations. (b) Regions of the Gaussian fitting in the direction of the line of sight. Each white line represents a fitting depth of 20 Mm at each pixel in the simulation. (c) Slit position of IRIS at the observation time under analysis. (d) The Mg II k spectrum from IRIS at 02:04 UT. (e) Doppler velocities calculated from both the observations (solid line) and simulations (dashed line). Here we show the results with the highest correlation coefficient of 0.71 for the corresponding times shown in the panel.}
    \label{fig:fig9}
\end{figure}

We further extract the velocity and density information from the simulation snapshots. And then we analyze the Doppler velocity along the line of sight and compare it with that from observations. The results are shown in Figure \ref{fig:fig9}. The Doppler velocity from the simulation is derived from the Gaussian fitting based on the distribution function of density squared versus velocity along the line of sight as shown by each white line in Figure \ref{fig:fig9}(b). Note that every white line is perpendicular to the IRIS slit and can be projected to one point on the slit. Then we can obtain the velocity distribution from simulation at these points along the slit. We discard the extremely large values compared with the maximum absolute value from observations and further make a spatial interpolation to the interval of observations. This is intended for calculating the cross-correlation between Doppler velocities from observations and simulations. Because of the lack of a lot of information about specific matters in the zero-$\beta$ simulation, we take such an averaged method in analyzing the Doppler velocities. For the observational result at a selected time, we compare it with each simulation snapshot. And the best result with a correlation coefficient of 0.71 is shown in Figure \ref{fig:fig9}(e). The time difference between the observation and simulation is about 6.5 minutes for this best matched case, which is acceptable considering the simplified MHD equations used in the simulation.

This jet event has already been studied by two different groups, as it was mentioned in Section \ref{sec:observations and method}. Different opinions were proposed regarding the trigger mechanism of the jet event. \citet{Yang_2020} did an NLFFF extrapolation to derive the main magnetic structure of this event. They confirmed a fan-spine magnetic topology and found two flux ropes in the extrapolated results. Also, they found a null point responsible for the reconnection site, which is the same as that we find in this work. In their opinion, the breakout-type reconnection caused by one of the flux ropes leads to the eruption of the jet. By contrast, \citet{Joshi_2020a} checked the magnetogram evolution prior to the occurrence of the event and analyzed the vector magnetic field map in comparison with an ideal MHD simulation. They found a flux rope along the polarity inversion line in the north and the flux rope migrates towards the south due to the photospheric motions, leading to the formation of a small parasitic bipole. And the rotational motion of the flux rope was verified by them through analysis on the IRIS spectra in their Figure 8(b). Though not so obvious, the rotational characteristics can also be exhibited in our Figure \ref{fig:fig9}(d). A strong extended bidirectional shifts at around $y=75$ and from $y=75$ blue shifts fade along the $y$ axis, while red shifts fade reverse the $y$ axis. They found that the bipole is also a bald patch region, above which a current sheet forms. Thus the twist of the flux rope is transferred into the jet through reconnection without direct eruptions. These two studies come into a consensus that the helical motion of the jet comes from the pre-existing twist in the flux rope. Nevertheless, the critical difference of these two studies lies in how the jet erupts, namely, how is the twist transferred into the jet. Such a difference could be ascribed to the different methods used in these two studies, since the study by \citet{Yang_2020} is based on the magnetic field extrapolation together with a high-resolution observation from NVST, while that of \citet{Joshi_2020a} focuses on the evolution of the magnetogram and comparisons with ideal MHD simulations. Neither of them performed dynamic simulations about this active region. Thus, we make a zero-$\beta$ MHD simulation in this work to probe how the jet is triggered. The Doppler velocity of our simulation is in accordance with IRIS observations from the aspect of correlation analysis. Specifically, our simulation shows that the flux rope lies low at the bald patch and the null point is located high in the corona. However, they are not far from each other when observed from the viewing angle of the spacecraft. In that sense, it is hard to distinguish their distinct features only according to the spectrum that is actually an integration along the line of sight. Therefore, pure analysis of observations might not lead to a solid conclusion. The combination of MHD simulations and observations is needed for learning the physical nature of solar eruptive events.

\section{Summary and Discussion} \label{sec:summary}

With the simultaneous observations from AIA and IRIS, we analyze the evolution of the jet in AR 12736 on 2019 March 22. We draw a time-slice map for the slit along the jet axis and find that the jet is twisting around its axis as propagating outwards at a speed of $\sim$ 300 km\ s$^{-1}$. The jet erupts with hot matter followed by cold matter. We suppose that they correspond to the hot reconnection outflows through a null point and the cold material from the flux rope. The helical motion comes from the untwisting of the flux rope. Furthermore, we utilize an NLFFF to check the topological structure of the magnetic field with the photospheric vector magnetic field from SOT/SP. A null point structure in the south and a flux rope in the north of a bald patch region are found to exist in the active region. Then, we adopt a zero-$\beta$ MHD simulation based on the results of the NLFFF extrapolation. It reveals that the jet may be produced by the magnetic reconnection at the null point. With the untwisting motion of the flux rope and slipping motion of its footpoint, the inner field lines of the flux rope encounter and then reconnect with the field lines originally situated at the null point. The reconnection outflows propagate along the spine line connected to the major negative polarity in the west region, which correspond to the bright and dark patterns in the AIA images.

We experiment with both the SDO/HMI magnetogram and SOT/SP magnetogram and the reconstruction of the NLFFF for further MHD simulations. An X-shaped current sheet was found in both cases; however, the flux rope can only be found in the SOT/SP case. Although we adopt the same magnetogram as that used in \citet{Yang_2020}, the magnetic structures obtained are different in the two cases. We attribute the different results to the usage of different extrapolation methods. Moreover, in our trials without extra anomalous resistivity, the rotational movement of the flux rope near the photosphere cannot be reproduced in the simulation. The resistivity is artificially added in the regions where the current density is large and thus the flux rope erupts with high resistivity and reveals untwisting motions. This implies that the rotational untwisting motion of the flux rope at the beginning comes from the reconnection in the flux rope. Furthermore, the ideal MHD simulations conducted by \citet{Wyper_2018} show that minority-polarity intrusions may add free energy into the field and can lead to both intermittent low-level reconnection and explosive, high-energy-release reconnection above these regions. Though the moving of the parasitic polarity is not considered in our data-constrained simulation, this may partly account for the impacts of the resistivity added in our experiments.

Our MHD simulation does not contain heat conduction and radiation process. Therefore, the physical parameters we obtain in the numerical simulations are not quantitatively accurate. Recently, \citet{Zhou_2022} conducted an MHD simulation of a sheared arcade configuration, in which a magnetic flux rope is formed and erupts through reconnection. The flux rope in their simulation rotates against its twist and transfers the twist into the surrounding field lines, which resembles our results. However, further studies are needed for clarifying the mechanism of the rotation of the flux rope leading to the reconnection between the flux rope and the spine structure connected to the remote side of the active region. In particular, we should focus on the role of the anomalous resistivity in driving the reconnection in the inner part of the flux rope and leading to its rotational motion around its axis and its upward motion in the chromosphere. We should also improve the data-constrained simulations, such as adding some specific magnetic structures and photospheric motions in order to fit the observations more closely. On the other hand, we also aim to carry out simulations with full MHD equations including heat conduction and radiative cooling to learn the physics more accurately in such jet events.

\begin{acknowledgements}
J.H.Z., Y.G. and M.D.D. are supported by the National Key Research and Development Program of China (2022YFF0503004, 2021YFA1600504, and 2020YFC2201201) and NSFC (11773016 and 11733003). We thank the two anonymous referees very much for their constructive suggestions. We deeply appreciate the free access to the magnetogram and observation data provided by SDO/AIA, SDO/HMI, SOT/SP and IRIS science teams. J.H.Z. thanks Ye Qiu and Ze Zhong for their help in the work. Computational resources were provided by the High Performance Computing Center (HPCC) at Nanjing University.
\end{acknowledgements}

\figsetplot

\begin{thebibliography}{}
\bibitem[{{Adams} {et~al.}(2014){Adams}, {Sterling}, {Moore}, \&
  {Gary}}]{Adams_2014}
{Adams}, M., {Sterling}, A.~C., {Moore}, R.~L., \& {Gary}, G.~A. 2014, \apj,
  783, 11, \dodoi{10.1088/0004-637X/783/1/11}

\bibitem[{{Aulanier} \& {Dud{\'\i}k}(2019)}]{Aulanier_2019}
{Aulanier}, G., \& {Dud{\'\i}k}, J. 2019, \aap, 621, A72,
  \dodoi{10.1051/0004-6361/201834221}

\bibitem[{{Baikie} {et~al.}(2022){Baikie}, {Sterling}, {Moore}, {Alexander},
  {Falconer}, {Savcheva}, \& {Savage}}]{Baikie_2022}
{Baikie}, T.~K., {Sterling}, A.~C., {Moore}, R.~L., {et~al.} 2022, \apj, 927,
  79, \dodoi{10.3847/1538-4357/ac473e}

\bibitem[{{Cirtain} {et~al.}(2007){Cirtain}, {Golub}, {Lundquist}, {van
  Ballegooijen}, {Savcheva}, {Shimojo}, {DeLuca}, {Tsuneta}, {Sakao}, {Reeves},
  {Weber}, {Kano}, {Narukage}, \& {Shibasaki}}]{Cirtain_2007}
{Cirtain}, J.~W., {Golub}, L., {Lundquist}, L., {et~al.} 2007, Science, 318,
  1580, \dodoi{10.1126/science.1147050}

\bibitem[{{De Pontieu} {et~al.}(2014){De Pontieu}, {Title}, {Lemen}, {Kushner},
  {Akin}, {Allard}, {Berger}, {Boerner}, {Cheung}, {Chou}, {Drake}, {Duncan},
  {Freeland}, {Heyman}, {Hoffman}, {Hurlburt}, {Lindgren}, {Mathur}, {Rehse},
  {Sabolish}, {Seguin}, {Schrijver}, {Tarbell}, {W{\"u}lser}, {Wolfson},
  {Yanari}, {Mudge}, {Nguyen-Phuc}, {Timmons}, {van Bezooijen}, {Weingrod},
  {Brookner}, {Butcher}, {Dougherty}, {Eder}, {Knagenhjelm}, {Larsen},
  {Mansir}, {Phan}, {Boyle}, {Cheimets}, {DeLuca}, {Golub}, {Gates}, {Hertz},
  {McKillop}, {Park}, {Perry}, {Podgorski}, {Reeves}, {Saar}, {Testa}, {Tian},
  {Weber}, {Dunn}, {Eccles}, {Jaeggli}, {Kankelborg}, {Mashburn}, {Pust},
  {Springer}, {Carvalho}, {Kleint}, {Marmie}, {Mazmanian}, {Pereira}, {Sawyer},
  {Strong}, {Worden}, {Carlsson}, {Hansteen}, {Leenaarts}, {Wiesmann},
  {Aloise}, {Chu}, {Bush}, {Scherrer}, {Brekke}, {Martinez-Sykora}, {Lites},
  {McIntosh}, {Uitenbroek}, {Okamoto}, {Gummin}, {Auker}, {Jerram}, {Pool}, \&
  {Waltham}}]{Pontieu_2014}
{De Pontieu}, B., {Title}, A.~M., {Lemen}, J.~R., {et~al.} 2014, \solphys, 289,
  2733, \dodoi{10.1007/s11207-014-0485-y}

\bibitem[{{Fang} {et~al.}(2014){Fang}, {Fan}, \& {McIntosh}}]{Fang_2014}
{Fang}, F., {Fan}, Y., \& {McIntosh}, S.~W. 2014, \apjl, 789, L19,
  \dodoi{10.1088/2041-8205/789/1/L19}

\bibitem[{{Farid} {et~al.}(2022){Farid}, {Savcheva}, {Tassav}, \&
  {Reeves}}]{Farid_2022}
{Farid}, S.~I., {Savcheva}, A., {Tassav}, S., \& {Reeves}, K.~K. 2022, \apj,
  938, 150, \dodoi{10.3847/1538-4357/ac8c2e}

\bibitem[{{Gonz{\'a}lez-Avil{\'e}s} {et~al.}(2020){Gonz{\'a}lez-Avil{\'e}s},
  {Guzm{\'a}n}, {Fedun}, \& {Verth}}]{Gonz_2020}
{Gonz{\'a}lez-Avil{\'e}s}, J.~J., {Guzm{\'a}n}, F.~S., {Fedun}, V., \& {Verth},
  G. 2020, \apj, 897, 153, \dodoi{10.3847/1538-4357/ab97b8}

\bibitem[{{Guo} {et~al.}(2013){Guo}, {D{\'e}moulin}, {Schmieder}, {Ding},
  {Vargas Dom{\'\i}nguez}, \& {Liu}}]{Guo_2013}
{Guo}, Y., {D{\'e}moulin}, P., {Schmieder}, B., {et~al.} 2013, \aap, 555, A19,
  \dodoi{10.1051/0004-6361/201321229}

\bibitem[{{Guo} {et~al.}(2016{\natexlab{a}}){Guo}, {Xia}, \&
  {Keppens}}]{Guo_2016a}
{Guo}, Y., {Xia}, C., \& {Keppens}, R. 2016{\natexlab{a}}, \apj, 828, 83,
  \dodoi{10.3847/0004-637X/828/2/83}

\bibitem[{{Guo} {et~al.}(2019){Guo}, {Xia}, {Keppens}, {Ding}, \&
  {Chen}}]{Guo_2019}
{Guo}, Y., {Xia}, C., {Keppens}, R., {Ding}, M.~D., \& {Chen}, P.~F. 2019,
  \apjl, 870, L21, \dodoi{10.3847/2041-8213/aafabf}

\bibitem[{{Guo} {et~al.}(2016{\natexlab{b}}){Guo}, {Xia}, {Keppens}, \&
  {Valori}}]{Guo_2016b}
{Guo}, Y., {Xia}, C., {Keppens}, R., \& {Valori}, G. 2016{\natexlab{b}}, \apj,
  828, 82, \dodoi{10.3847/0004-637X/828/2/82}

\bibitem[{{Ichimoto} {et~al.}(2008){Ichimoto}, {Lites}, {Elmore}, {Suematsu},
  {Tsuneta}, {Katsukawa}, {Shimizu}, {Shine}, {Tarbell}, {Title}, {Kiyohara},
  {Shinoda}, {Card}, {Lecinski}, {Streander}, {Nakagiri}, {Miyashita},
  {Noguchi}, {Hoffmann}, \& {Cruz}}]{SOT2008-4}
{Ichimoto}, K., {Lites}, B., {Elmore}, D., {et~al.} 2008, \solphys, 249, 233,
  \dodoi{10.1007/s11207-008-9169-9}

\bibitem[{{Jiang} {et~al.}(2022){Jiang}, {Feng}, {Guo}, \& {Hu}}]{Jiang_2022}
{Jiang}, C., {Feng}, X., {Guo}, Y., \& {Hu}, Q. 2022, The Innovation, 3,
  100236, \dodoi{10.1016/j.xinn.2022.100236}

\bibitem[{{Jiang} {et~al.}(2016){Jiang}, {Wu}, {Feng}, \& {Hu}}]{Jiang_2016}
{Jiang}, C., {Wu}, S.~T., {Feng}, X., \& {Hu}, Q. 2016, Nature Communications,
  7, 11522, \dodoi{10.1038/ncomms11522}

\bibitem[{{Joshi} {et~al.}(2020{\natexlab{a}}){Joshi}, {Schmieder}, {Aulanier},
  {Bommier}, \& {Chandra}}]{Joshi_2020a}
{Joshi}, R., {Schmieder}, B., {Aulanier}, G., {Bommier}, V., \& {Chandra}, R.
  2020{\natexlab{a}}, \aap, 642, A169, \dodoi{10.1051/0004-6361/202038562}

\bibitem[{{Joshi} {et~al.}(2021{\natexlab{a}}){Joshi}, {Schmieder}, {Heinzel},
  {Tomin}, {Chandra}, \& {Vilmer}}]{Joshi_2021b}
{Joshi}, R., {Schmieder}, B., {Heinzel}, P., {et~al.} 2021{\natexlab{a}}, \aap,
  654, A31, \dodoi{10.1051/0004-6361/202141172}

\bibitem[{{Joshi} {et~al.}(2021{\natexlab{b}}){Joshi}, {Schmieder}, {Tei},
  {Aulanier}, {L{\"o}rin{\v{c}}{\'\i}k}, {Chandra}, \& {Heinzel}}]{Joshi_2021a}
{Joshi}, R., {Schmieder}, B., {Tei}, A., {et~al.} 2021{\natexlab{b}}, \aap,
  645, A80, \dodoi{10.1051/0004-6361/202039229}

\bibitem[{{Joshi} {et~al.}(2020{\natexlab{b}}){Joshi}, {Wang}, {Chandra},
  {Zhang}, {Liu}, \& {Li}}]{Joshi_2020b}
{Joshi}, R., {Wang}, Y., {Chandra}, R., {et~al.} 2020{\natexlab{b}}, \apj, 901,
  94, \dodoi{10.3847/1538-4357/abaf5a}

\bibitem[{{Keppens} {et~al.}(2012){Keppens}, {Meliani}, {van Marle}, {Delmont},
  {Vlasis}, \& {van der Holst}}]{Keppens_2012}
{Keppens}, R., {Meliani}, Z., {van Marle}, A.~J., {et~al.} 2012, Journal of
  Computational Physics, 231, 718, \dodoi{10.1016/j.jcp.2011.01.020}

\bibitem[{{Keppens} {et~al.}(2003){Keppens}, {Nool}, {T{\'o}th}, \&
  {Goedbloed}}]{Keppens_2003}
{Keppens}, R., {Nool}, M., {T{\'o}th}, G., \& {Goedbloed}, J.~P. 2003, Computer
  Physics Communications, 153, 317, \dodoi{10.1016/S0010-4655(03)00139-5}

\bibitem[{{Kosugi} {et~al.}(2007){Kosugi}, {Matsuzaki}, {Sakao}, {Shimizu},
  {Sone}, {Tachikawa}, {Hashimoto}, {Minesugi}, {Ohnishi}, {Yamada}, {Tsuneta},
  {Hara}, {Ichimoto}, {Suematsu}, {Shimojo}, {Watanabe}, {Shimada}, {Davis},
  {Hill}, {Owens}, {Title}, {Culhane}, {Harra}, {Doschek}, \&
  {Golub}}]{Hinode_2007}
{Kosugi}, T., {Matsuzaki}, K., {Sakao}, T., {et~al.} 2007, \solphys, 243, 3,
  \dodoi{10.1007/s11207-007-9014-6}

\bibitem[{{Lemen} {et~al.}(2012){Lemen}, {Title}, {Akin}, {Boerner}, {Chou},
  {Drake}, {Duncan}, {Edwards}, {Friedlaender}, {Heyman}, {Hurlburt}, {Katz},
  {Kushner}, {Levay}, {Lindgren}, {Mathur}, {McFeaters}, {Mitchell}, {Rehse},
  {Schrijver}, {Springer}, {Stern}, {Tarbell}, {Wuelser}, {Wolfson}, {Yanari},
  {Bookbinder}, {Cheimets}, {Caldwell}, {Deluca}, {Gates}, {Golub}, {Park},
  {Podgorski}, {Bush}, {Scherrer}, {Gummin}, {Smith}, {Auker}, {Jerram},
  {Pool}, {Soufli}, {Windt}, {Beardsley}, {Clapp}, {Lang}, \&
  {Waltham}}]{lemen_2012}
{Lemen}, J.~R., {Title}, A.~M., {Akin}, D.~J., {et~al.} 2012, \solphys, 275,
  17, \dodoi{10.1007/s11207-011-9776-8}

\bibitem[{{Lites} {et~al.}(2013){Lites}, {Akin}, {Card}, {Cruz}, {Duncan},
  {Edwards}, {Elmore}, {Hoffmann}, {Katsukawa}, {Katz}, {Kubo}, {Ichimoto},
  {Shimizu}, {Shine}, {Streander}, {Suematsu}, {Tarbell}, {Title}, \&
  {Tsuneta}}]{SP2013}
{Lites}, B.~W., {Akin}, D.~L., {Card}, G., {et~al.} 2013, \solphys, 283, 579,
  \dodoi{10.1007/s11207-012-0206-3}

\bibitem[{{Liu} {et~al.}(2011{\natexlab{a}}){Liu}, {Deng}, {Liu},
  {Ugarte-Urra}, {Wang}, \& {Wang}}]{LiuC_2011}
{Liu}, C., {Deng}, N., {Liu}, R., {et~al.} 2011{\natexlab{a}}, \apjl, 735, L18,
  \dodoi{10.1088/2041-8205/735/1/L18}

\bibitem[{{Liu} {et~al.}(2011{\natexlab{b}}){Liu}, {Berger}, {Title},
  {Tarbell}, \& {Low}}]{Liu_2011}
{Liu}, W., {Berger}, T.~E., {Title}, A.~M., {Tarbell}, T.~D., \& {Low}, B.~C.
  2011{\natexlab{b}}, \apj, 728, 103, \dodoi{10.1088/0004-637X/728/2/103}

\bibitem[{{Metcalf} {et~al.}(2008){Metcalf}, {De Rosa}, {Schrijver}, {Barnes},
  {van Ballegooijen}, {Wiegelmann}, {Wheatland}, {Valori}, \&
  {McTtiernan}}]{Metcalf_2008}
{Metcalf}, T.~R., {De Rosa}, M.~L., {Schrijver}, C.~J., {et~al.} 2008,
  \solphys, 247, 269, \dodoi{10.1007/s11207-007-9110-7}

\bibitem[{{Moore} {et~al.}(2010){Moore}, {Cirtain}, {Sterling}, \&
  {Falconer}}]{Moore_2010}
{Moore}, R.~L., {Cirtain}, J.~W., {Sterling}, A.~C., \& {Falconer}, D.~A. 2010,
  \apj, 720, 757, \dodoi{10.1088/0004-637X/720/1/757}

\bibitem[{{Nistic{\`o}} {et~al.}(2009){Nistic{\`o}}, {Bothmer}, {Patsourakos},
  \& {Zimbardo}}]{Nistico_2009}
{Nistic{\`o}}, G., {Bothmer}, V., {Patsourakos}, S., \& {Zimbardo}, G. 2009,
  \solphys, 259, 87, \dodoi{10.1007/s11207-009-9424-8}

\bibitem[{{Nitta} {et~al.}(2006){Nitta}, {Reames}, {De Rosa}, {Liu}, {Yashiro},
  \& {Gopalswamy}}]{Nitta_2006}
{Nitta}, N.~V., {Reames}, D.~V., {De Rosa}, M.~L., {et~al.} 2006, \apj, 650,
  438, \dodoi{10.1086/507442}

\bibitem[{{Paraschiv} {et~al.}(2010){Paraschiv}, {Lacatus}, {Badescu}, {Lupu},
  {Simon}, {Sandu}, {Mierla}, \& {Rusu}}]{Paraschiv_2010}
{Paraschiv}, A.~R., {Lacatus}, D.~A., {Badescu}, T., {et~al.} 2010, \solphys,
  264, 365, \dodoi{10.1007/s11207-010-9584-6}

\bibitem[{{Pariat} {et~al.}(2010){Pariat}, {Antiochos}, \&
  {DeVore}}]{Pariat_2010}
{Pariat}, E., {Antiochos}, S.~K., \& {DeVore}, C.~R. 2010, \apj, 714, 1762,
  \dodoi{10.1088/0004-637X/714/2/1762}

\bibitem[{{Pariat} {et~al.}(2015){Pariat}, {Dalmasse}, {DeVore}, {Antiochos},
  \& {Karpen}}]{Pariat_2015}
{Pariat}, E., {Dalmasse}, K., {DeVore}, C.~R., {Antiochos}, S.~K., \& {Karpen},
  J.~T. 2015, \aap, 573, A130, \dodoi{10.1051/0004-6361/201424209}

\bibitem[{{Parnell} {et~al.}(1996){Parnell}, {Smith}, {Neukirch}, \&
  {Priest}}]{Parnell_1996}
{Parnell}, C.~E., {Smith}, J.~M., {Neukirch}, T., \& {Priest}, E.~R. 1996,
  Physics of Plasmas, 3, 759, \dodoi{10.1063/1.871810}

\bibitem[{{Patsourakos} {et~al.}(2008){Patsourakos}, {Pariat}, {Vourlidas},
  {Antiochos}, \& {Wuelser}}]{Patsourakos_2008}
{Patsourakos}, S., {Pariat}, E., {Vourlidas}, A., {Antiochos}, S.~K., \&
  {Wuelser}, J.~P. 2008, \apjl, 680, L73, \dodoi{10.1086/589769}

\bibitem[{{Pesnell} {et~al.}(2012){Pesnell}, {Thompson}, \&
  {Chamberlin}}]{SDO2012}
{Pesnell}, W.~D., {Thompson}, B.~J., \& {Chamberlin}, P.~C. 2012, \solphys,
  275, 3, \dodoi{10.1007/s11207-011-9841-3}

\bibitem[{{Porth} {et~al.}(2014){Porth}, {Xia}, {Hendrix}, {Moschou}, \&
  {Keppens}}]{Porth_2014}
{Porth}, O., {Xia}, C., {Hendrix}, T., {Moschou}, S.~P., \& {Keppens}, R. 2014,
  \apjs, 214, 4, \dodoi{10.1088/0067-0049/214/1/4}

\bibitem[{{Priest} \& {Titov}(1996)}]{Priest_1996}
{Priest}, E.~R., \& {Titov}, V.~S. 1996, Philosophical Transactions of the
  Royal Society of London Series A, 354, 2951, \dodoi{10.1098/rsta.1996.0136}

\bibitem[{{Raouafi} {et~al.}(2016){Raouafi}, {Patsourakos}, {Pariat}, {Young},
  {Sterling}, {Savcheva}, {Shimojo}, {Moreno-Insertis}, {DeVore}, {Archontis},
  {T{\"o}r{\"o}k}, {Mason}, {Curdt}, {Meyer}, {Dalmasse}, \&
  {Matsui}}]{Raouafi_2016}
{Raouafi}, N.~E., {Patsourakos}, S., {Pariat}, E., {et~al.} 2016, \ssr, 201, 1,
  \dodoi{10.1007/s11214-016-0260-5}

\bibitem[{{Scherrer} {et~al.}(2012){Scherrer}, {Schou}, {Bush}, {Kosovichev},
  {Bogart}, {Hoeksema}, {Liu}, {Duvall}, {Zhao}, {Title}, {Schrijver},
  {Tarbell}, \& {Tomczyk}}]{Scherrer_2012}
{Scherrer}, P.~H., {Schou}, J., {Bush}, R.~I., {et~al.} 2012, \solphys, 275,
  207, \dodoi{10.1007/s11207-011-9834-2}

\bibitem[{{Schmieder}(2022)}]{Schmieder_2022a}
{Schmieder}, B. 2022, Frontiers in Astronomy and Space Sciences, 9, 820183,
  \dodoi{10.3389/fspas.2022.820183}

\bibitem[{{Schmieder} {et~al.}(2022){Schmieder}, {Joshi}, \&
  {Chandra}}]{Schmieder_2022b}
{Schmieder}, B., {Joshi}, R., \& {Chandra}, R. 2022, Advances in Space
  Research, 70, 1580, \dodoi{10.1016/j.asr.2021.12.013}

\bibitem[{{Schmieder} {et~al.}(2013){Schmieder}, {Guo}, {Moreno-Insertis},
  {Aulanier}, {Yelles Chaouche}, {Nishizuka}, {Harra}, {Thalmann}, {Vargas
  Dominguez}, \& {Liu}}]{Schmieder_2013}
{Schmieder}, B., {Guo}, Y., {Moreno-Insertis}, F., {et~al.} 2013, \aap, 559,
  A1, \dodoi{10.1051/0004-6361/201322181}

\bibitem[{{Schou} {et~al.}(2012){Schou}, {Scherrer}, {Bush}, {Wachter},
  {Couvidat}, {Rabello-Soares}, {Bogart}, {Hoeksema}, {Liu}, {Duvall}, {Akin},
  {Allard}, {Miles}, {Rairden}, {Shine}, {Tarbell}, {Title}, {Wolfson},
  {Elmore}, {Norton}, \& {Tomczyk}}]{Schou_2012}
{Schou}, J., {Scherrer}, P.~H., {Bush}, R.~I., {et~al.} 2012, \solphys, 275,
  229, \dodoi{10.1007/s11207-011-9842-2}

\bibitem[{{Schrijver} {et~al.}(2006){Schrijver}, {De Rosa}, {Metcalf}, {Liu},
  {McTiernan}, {R{\'e}gnier}, {Valori}, {Wheatland}, \&
  {Wiegelmann}}]{Schrijver_2006}
{Schrijver}, C.~J., {De Rosa}, M.~L., {Metcalf}, T.~R., {et~al.} 2006,
  \solphys, 235, 161, \dodoi{10.1007/s11207-006-0068-7}

\bibitem[{{Sheeley} \& {Wang}(2007)}]{Sheeley_2007}
{Sheeley}, N.~R., J., \& {Wang}, Y.~M. 2007, \apj, 655, 1142,
  \dodoi{10.1086/510323}

\bibitem[{{Shibata} {et~al.}(1994){Shibata}, {Nitta}, {Strong}, {Matsumoto},
  {Yokoyama}, {Hirayama}, {Hudson}, \& {Ogawara}}]{Shibata_1994}
{Shibata}, K., {Nitta}, N., {Strong}, K.~T., {et~al.} 1994, \apjl, 431, L51,
  \dodoi{10.1086/187470}

\bibitem[{{Shibata} {et~al.}(1992){Shibata}, {Ishido}, {Acton}, {Strong},
  {Hirayama}, {Uchida}, {McAllister}, {Matsumoto}, {Tsuneta}, {Shimizu},
  {Hara}, {Sakurai}, {Ichimoto}, {Nishino}, \& {Ogawara}}]{Shibata_1992}
{Shibata}, K., {Ishido}, Y., {Acton}, L.~W., {et~al.} 1992, \pasj, 44, L173

\bibitem[{{Shimizu} {et~al.}(2008){Shimizu}, {Nagata}, {Tsuneta}, {Tarbell},
  {Edwards}, {Shine}, {Hoffmann}, {Thomas}, {Sour}, {Rehse}, {Ito},
  {Kashiwagi}, {Tabata}, {Kodeki}, {Nagase}, {Matsuzaki}, {Kobayashi},
  {Ichimoto}, \& {Suematsu}}]{SOT2008-2}
{Shimizu}, T., {Nagata}, S., {Tsuneta}, S., {et~al.} 2008, \solphys, 249, 221,
  \dodoi{10.1007/s11207-007-9053-z}

\bibitem[{{Sterling} {et~al.}(2022){Sterling}, {Moore}, \&
  {Panesar}}]{Sterling_2022}
{Sterling}, A.~C., {Moore}, R.~L., \& {Panesar}, N.~K. 2022, \apj, 927, 127,
  \dodoi{10.3847/1538-4357/ac473f}

\bibitem[{{Suematsu} {et~al.}(2008){Suematsu}, {Tsuneta}, {Ichimoto},
  {Shimizu}, {Otsubo}, {Katsukawa}, {Nakagiri}, {Noguchi}, {Tamura}, {Kato},
  {Hara}, {Kubo}, {Mikami}, {Saito}, {Matsushita}, {Kawaguchi}, {Nakaoji},
  {Nagae}, {Shimada}, {Takeyama}, \& {Yamamuro}}]{SOT2008-1}
{Suematsu}, Y., {Tsuneta}, S., {Ichimoto}, K., {et~al.} 2008, \solphys, 249,
  197, \dodoi{10.1007/s11207-008-9129-4}

\bibitem[{{Titov} {et~al.}(1993){Titov}, {Priest}, \& {Demoulin}}]{Titov_1993}
{Titov}, V.~S., {Priest}, E.~R., \& {Demoulin}, P. 1993, \aap, 276, 564

\bibitem[{{Tsuneta} {et~al.}(2008){Tsuneta}, {Ichimoto}, {Katsukawa}, {Nagata},
  {Otsubo}, {Shimizu}, {Suematsu}, {Nakagiri}, {Noguchi}, {Tarbell}, {Title},
  {Shine}, {Rosenberg}, {Hoffmann}, {Jurcevich}, {Kushner}, {Levay}, {Lites},
  {Elmore}, {Matsushita}, {Kawaguchi}, {Saito}, {Mikami}, {Hill}, \&
  {Owens}}]{SOT2008-3}
{Tsuneta}, S., {Ichimoto}, K., {Katsukawa}, Y., {et~al.} 2008, \solphys, 249,
  167, \dodoi{10.1007/s11207-008-9174-z}

\bibitem[{{Wiegelmann} {et~al.}(2006){Wiegelmann}, {Inhester}, \&
  {Sakurai}}]{Wiegelmann_2006}
{Wiegelmann}, T., {Inhester}, B., \& {Sakurai}, T. 2006, \solphys, 233, 215,
  \dodoi{10.1007/s11207-006-2092-z}

\bibitem[{{Wiegelmann} \& {Sakurai}(2021)}]{Wiegelmann_2021}
{Wiegelmann}, T., \& {Sakurai}, T. 2021, Living Reviews in Solar Physics, 18,
  1, \dodoi{10.1007/s41116-020-00027-4}

\bibitem[{{Wyper} {et~al.}(2017){Wyper}, {Antiochos}, \& {DeVore}}]{Wyper_2017}
{Wyper}, P.~F., {Antiochos}, S.~K., \& {DeVore}, C.~R. 2017, \nat, 544, 452,
  \dodoi{10.1038/nature22050}

\bibitem[{{Wyper} {et~al.}(2019){Wyper}, {DeVore}, \& {Antiochos}}]{Wyper_2019}
{Wyper}, P.~F., {DeVore}, C.~R., \& {Antiochos}, S.~K. 2019, \mnras, 490, 3679,
  \dodoi{10.1093/mnras/stz2674}

\bibitem[{{Wyper} {et~al.}(2018){Wyper}, {DeVore}, {Karpen}, {Antiochos}, \&
  {Yeates}}]{Wyper_2018}
{Wyper}, P.~F., {DeVore}, C.~R., {Karpen}, J.~T., {Antiochos}, S.~K., \&
  {Yeates}, A.~R. 2018, \apj, 864, 165, \dodoi{10.3847/1538-4357/aad9f7}

\bibitem[{{Xia} {et~al.}(2018){Xia}, {Teunissen}, {El Mellah}, {Chan{\'e}}, \&
  {Keppens}}]{Xia_2018}
{Xia}, C., {Teunissen}, J., {El Mellah}, I., {Chan{\'e}}, E., \& {Keppens}, R.
  2018, \apjs, 234, 30, \dodoi{10.3847/1538-4365/aaa6c8}

\bibitem[{{Yang} {et~al.}(2020){Yang}, {Zhang}, {Xu}, {Zhang}, {Zhong}, \&
  {Guo}}]{Yang_2020}
{Yang}, S., {Zhang}, Q., {Xu}, Z., {et~al.} 2020, \apj, 898, 101,
  \dodoi{10.3847/1538-4357/ab9ac7}

\bibitem[{{Yu} {et~al.}(2014){Yu}, {Jackson}, {Buffington}, {Hick}, {Shimojo},
  \& {Sako}}]{Yu_2014}
{Yu}, H.~S., {Jackson}, B.~V., {Buffington}, A., {et~al.} 2014, \apj, 784, 166,
  \dodoi{10.1088/0004-637X/784/2/166}

\bibitem[{{Zhang} {et~al.}(2022){Zhang}, {Zhang}, {Dai}, {Li}, \&
  {Ji}}]{Zhang_2022}
{Zhang}, Y., {Zhang}, Q., {Dai}, J., {Li}, D., \& {Ji}, H. 2022, \solphys, 297,
  138, \dodoi{10.1007/s11207-022-02072-8}

\bibitem[{{Zhou} {et~al.}(2022){Zhou}, {Jiang}, {Liu}, {Wang}, {Liu}, \&
  {Cui}}]{Zhou_2022}
{Zhou}, Z., {Jiang}, C., {Liu}, R., {et~al.} 2022, \apjl, 927, L14,
  \dodoi{10.3847/2041-8213/ac5740}

\bibitem[{{Zhu} {et~al.}(2017){Zhu}, {Wang}, {Cheng}, \& {Huang}}]{Zhu_2017}
{Zhu}, X., {Wang}, H., {Cheng}, X., \& {Huang}, C. 2017, \apjl, 844, L20,
  \dodoi{10.3847/2041-8213/aa8033}

\bibitem[{{Zhu} {et~al.}(2016){Zhu}, {Wang}, {Du}, \& {He}}]{Zhu_2016}
{Zhu}, X., {Wang}, H., {Du}, Z., \& {He}, H. 2016, \apj, 826, 51,
  \dodoi{10.3847/0004-637X/826/1/51}

\end{thebibliography}
\end{document}